\pdfoutput=1
\documentclass[sigconf]{acmart}
\usepackage{booktabs}

\usepackage{booktabs}
\usepackage{todonotes}
\usepackage{enumitem}
\usepackage{multirow}
\usepackage{graphicx}

\definecolor{darkgreen}{rgb}{0,0.5,0}
\definecolor{orange}{rgb}{1,0.5,0}
\definecolor{teal}{rgb}{0,0.5,0.5}
\definecolor{darkpurple}{rgb}{0.5, 0, 0.5}
\definecolor{LGorang}{rgb}{0.81,0.43,0.24}

\AtBeginDocument{%
  \providecommand\BibTeX{{%
    \normalfont B\kern-0.5em{\scshape i\kern-0.25em b}\kern-0.8em\TeX}}}

\copyrightyear{2023}
\acmYear{2023}
\setcopyright{rightsretained}
\acmConference[ASSETS '23]{The 25th International ACM SIGACCESS Conference
on Computers and Accessibility}{October 22--25, 2023}{New York, NY, USA}
\acmBooktitle{The 25th International ACM SIGACCESS Conference on Computers
and Accessibility (ASSETS '23), October 22--25, 2023, New York, NY, USA}
\acmDOI{10.1145/3597638.3608425}
\acmISBN{979-8-4007-0220-4/23/10}

\acmPrice{15.00}
\acmISBN{978-1-4503-XXXX-X/18/06}

\begin{document}

\title{Exploring Community-Driven Descriptions for Making Livestreams Accessible}

\author{Daniel Killough}
\affiliation{%
  \institution{The University of Texas at Austin \\ Department of Computer Science}
  \city{Austin}
  \state{Texas}
  \country{USA}
}
\email{contact@dkillough.com}

\author{Amy Pavel}
\affiliation{%
  \institution{The University of Texas at Austin \\ Department of Computer Science}
  \city{Austin}
  \state{Texas}
  \country{USA}
}
\email{apavel@cs.utexas.edu}

\renewcommand{\shortauthors}{Killough and Pavel}

\begin{abstract}
People watch livestreams to connect with others and learn about their hobbies. Livestreams feature multiple visual streams including the main video, webcams, on-screen overlays, and chat, all of which are inaccessible to livestream viewers with visual impairments. While prior work explores creating audio descriptions for recorded videos, live videos present new challenges: authoring descriptions in real-time, describing domain-specific content, and prioritizing which complex visual information to describe. We explore inviting livestream community members who are domain experts to provide live descriptions. We first conducted a study with 18 sighted livestream community members authoring descriptions for livestreams using three different description methods: live descriptions using text, live descriptions using speech, and asynchronous descriptions using text. We then conducted a study with 9 livestream community members with visual impairments, who shared their current strategies and challenges for watching livestreams and provided feedback on the community-written descriptions. We conclude with implications for improving the accessibility of livestreams.
\end{abstract}




\maketitle

\section{Introduction}
\begin{figure*}
    \centering
    \includegraphics[width=7in]{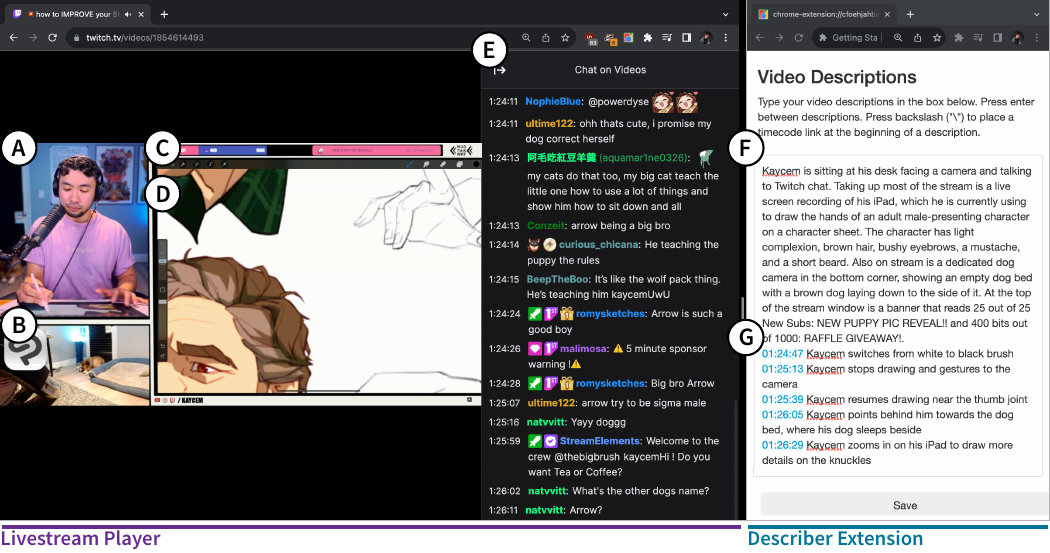}
    \caption{The \textit{Livestream Player} (left) features the livestream (A-D) and audience live chat (E). The livestream includes webcams for the streamer (A) and a dog (B), an overlay with status indicators (C), and the main video displaying a screenshare of a creative application (D).  The \textit{Describer Extension} (right) enables describers to input text descriptions while the livestream plays. Pressing the backslash key while a Livestream Player window is open inserts the current video timecode into the textbox (F). Clicking a timecode (G) seeks the Livestream Player video to the corresponding playback time. Source: Twitch livestream \textit{How to IMPROVE your SKILLS QUICKLY! Character Design Bootcamp \#2 Day 06/30 !bootcamp !youtube !resources} by Kaycem~\protect{\cite{kaycem}}.}
    \Description[A figure depicting the description input Chrome Extension.]{Figure 1: A figure depicting two Google Chrome windows side-by side. On the left Google Chrome window is a twitch.tv stream marked with the letter A, featuring a man drawing a male-presenting character on a character sheet. The man is holding an Apple Pencil and talking into a microphone on a boom arm. He is looking down towards an iPad laying flat on his desk. A fullscreen view of the iPad is seen to the right of the man, which shows a line art drawing of a squinting man with wavy brown hair, busy eyebrows, and a green flannel collared shirt. To the right of the stream window is a full column view of a chatbox with many multicolored usernames saying different chat messages with timestamps ranging from 1:24:11 to 1:26:11 as they progress downwards. The most recent messages ask, ``What’s the other dog’s name? Arrow?''. On the right Google Chrome window is a text entry box with the title ``Video Descriptions''. The subtitle reads ``Type your video descriptions in the box below. Press enter between descriptions. Press backslash (`\textbackslash') to place a timecode link at the beginning of a description.'' Within the text entry box is a series of visual descriptions marked with the letter F describing the content happening in the Twitch stream from the left Google Chrome window. The first 12 lines of the text box are a contextual description of the Twitch stream’s scene. The next 5 lines start with a timestamp starting at 1:24:47 until 1:26:29 with text following, each describing some action happening in the stream at that timestamp. At the bottom of the text box is a full-width gray Save button.}
    \label{fig:chromeextension}
\end{figure*}

Live videos (i.e. ``livestreams'') shared on streaming services like YouTube~\cite{youtubelive}, Facebook~\cite{fblive}, and Twitch~\cite{twitch} are becoming increasingly popular.
People authoring livestreams (i.e. ``streamers'') broadcast long activities in real-time, such as playing games~\cite{hamilton2014streaming}, producing art~\cite{fraser2019sharing}, leading educational exercises~\cite{chen2021afraid}, coding~\cite{faas2018watch}, or exploring the outdoors~\cite{lu2019vicariously}.
During a livestream, the streamer and audience synchronously interact with each other via the streamer's live video --- including webcams (Figure~\ref{fig:chromeextension}A-B), on-screen overlays (Figure~\ref{fig:chromeextension}C), and main activity video (Figure~\ref{fig:chromeextension}D) --- as well as with the audience live chat (Figure~\ref{fig:chromeextension}E). 
Real-time interaction provides viewers a chance to suggest next steps, ask for clarifications, and discuss reactions. 
The format of livestreams thus enables online communities to form around shared experiences~\cite{hamilton2014streaming}, and even extend offline~\cite{sheng2020virtual}. However, the rich visual content of livestreams that affords community building is not accessible to people with visual impairments.

To make recorded videos accessible, people add narration of the important visual content in the video, i.e. \textit{audio descriptions}.
Prior work explored how to create audio descriptions for recorded videos such as films~\cite{branje2012livedescribe,adpguidelines,snyder2005audio}, user-generated videos~\cite{liu2021what,liu2022crossa11y,pavel2020rescribe,youdescribe,wang2021toward}, slide presentations~\cite{peng2021slidecho,peng2021say} and GIFs~\cite{gleason2020making} by providing computational description support~\cite{pavel2020rescribe,liu2022crossa11y,peng2021slidecho,wang2021toward,yuksel2020human} and proposing what to describe for specific video types (\textit{e.g.}, GIFs~\cite{gleason2020making}, films~\cite{snyder2005audio}).
Previous work has not yet explored technology to support live descriptions or description preferences for livestream-specific content (\textit{e.g.}, long expert streams). 
Jun et al. investigated the accessibility of livestreaming for \textit{streamers} with visual impairments~\cite{jun2021exploring}. Streamers have accessibility needs that overlap with those of viewers (\textit{e.g.}, accessing chat), yet it remains unclear how to make live videos accessible. 
While online services provide on-demand live descriptions~\cite{airaio,bemyeyes}, streams are challenging to understand for people without domain familiarity due to complex visual content (\textit{e.g.}, multiplayer gameplay, expert software). 
Following the success of community-driven efforts to make videos accessible for d/Deaf and Hard of Hearing audience members via fansubbing~\cite{lee2011participatory} or community captions~\cite{huang2017leveraging}, and drawing on community  sourcing~\cite{heimerl2012communitysourcing,kim2015learnersourcing}, we invite sighted livestream viewers familiar with the livestream content to make livestreams non-visually accessible.

We present two studies exploring the feasibility of community-driven livestream accessibility. We first invited 18 sighted livestream community members to author descriptions of livestreams in domains that they were familiar with to compare three description approaches: live description using voice, live description using text, and asynchronous description using text. We also interviewed 9 livestream community members with visual impairments to share their current livestream viewing practices and challenges and provide feedback on the descriptions written by community members.

Overall, sighted community members generated descriptions that increased the accessibility of livestreams using all description methods. While sighted community members found it more challenging to provide live rather than asynchronous descriptions, they adapted several strategies to successfully create live descriptions including: describing during the streamer's narration, using domain-specific terms to quickly author descriptions (\textit{e.g.}, ``Up-B'' to describe a character's special attack in a game), and primarily describing updates due to individual actions (i.e. play-by-plays) rather than the scene as a whole. For providing live descriptions, community members differed in their preference for text vs. voice for description input.
However, community members provided significantly more descriptions and description words per video minute using voice input than using text input for live descriptions. 

Community members with visual impairments reported accessibility issues with consuming livestreams due to the platform's interface and the livestream content itself. 
Though most community members with visual impairments interviewed use YouTube instead of Twitch to avoid platform accessibility issues, livestream content remained inaccessible. 
Community members reported that the streamers' speech often diverged from describing their actions (\textit{e.g.}, telling a story while creating an art piece) and used frequent visual references to other parts of the video that were difficult to understand (\textit{e.g.}, reacting to an unknown chat message or referring an on-camera event). Viewers found community-written descriptions to be valuable in understanding the video as they filled in gaps left by the speaker. Viewers also suggested improvements for future descriptions, such as providing adjustable preferences on the expertise level, level of detail, and amount of overlap with the audio channel.
We conclude with directions for future systems aiming to make livestreams accessible.

In summary, we contribute:
\begin{itemize}
    \item An exploratory study with livestream community members providing descriptions of live video
    \item Interviews with livestream viewers with visual impairments sharing current strategies and challenges for watching live-streams
    \item Description preferences from livestream viewers with visual impairments derived from a co-watching exercise and feedback on community-written descriptions
\end{itemize}

\section{Background}
Our work builds upon prior work in video, livestreaming accessibility, and crowdsourcing for accessibility.

\subsection{Video Accessibility}
To make videos accessible to people with visual impairments, professionals traditionally create \textit{audio descriptions}, or narrations of the important visual content in a scene that cannot be understood from the audio alone~\cite{adp}. While audio descriptions increasingly exist for films and TV, they rarely exist for user-generated content. Prior work developed tools to make authoring audio descriptions easier by generating them automatically~\cite{wang2021toward}, or aiding novices in authoring audio descriptions~\cite{pavel2020rescribe,liu2022crossa11y,branje2012livedescribe,youdescribe,natalie2020viscene,natalie2023supporting,yuksel2020human}. For example, prior work helped novices edit their descriptions to fit into times without narration~\cite{pavel2020rescribe}, identify parts of the video likely to be inaccessible~\cite{liu2022crossa11y}, host their descriptions~\cite{youdescribe}, gain feedback on their descriptions~\cite{natalie2023supporting}, and locate silences~\cite{branje2012livedescribe,pavel2020rescribe}. 
These systems all process recorded videos rather than live videos, such that they are not suitable for livestreams. Such video accessibility work also explored generally understandable visual content rather than the domain-specific visual content present in livestreams. We explore how community member familiar with the domain may be able to provide descriptions for live rather than recorded videos.

While audio descriptions typically occur within gaps in video narration~\cite{pavel2020rescribe,branje2012livedescribe}, adequate gaps do not always occur (\textit{e.g.}, for short videos~\cite{gleason2020making}, or videos with frequent speech~\cite{pavel2020rescribe,peng2021slidecho}). To address this time constraint, prior work used rich audio to convey video themes~\cite{gleason2020making}, and provided users control over how often or when to pause a video to receive additional descriptions~\cite{pavel2020rescribe,peng2021slidecho}. Live video presents new time constraints for describers aiming to describe content as it happens, as well as for listeners aiming to keep up with the video pace. We investigate the feasibility of producing and consuming descriptions under such time constraints.

\subsection{Livestreams and Accessibility}
Livestreaming, broadcasting live video over the internet, has grown over recent decades with increased internet speeds and a broad selection of platforms (\textit{e.g.}, justin.tv now Twitch, Facebook Live, YouTube Live, TikTok LIVE).
We discuss livestream features common on platforms such as Twitch and YouTube Live to reflect on implications for accessibility for viewers with visual impairments:

\textbf{Long, real-time broadcasts}: As livestreams are broadcast in real time, streams are often unedited and occur over long durations (\textit{e.g.}, up to 5 hours or more~\cite{lu2019vicariously}). Compared to edited, recorded videos, livestreams activate communities around watching the content in real time~\cite{smith2013live}, engage viewers with one another for more time~\cite{hamilton2014streaming}, and enable viewers to gain depth in the streamed activity (\textit{e.g.} watching a game rather than highlights; seeing an artist work instead of explain the high level steps). While viewers may watch the livestream for a long time (\textit{e.g.}, 5 hours~\cite{lu2019vicariously}), they may join in the middle of the stream and need to ``catch up''~\cite{yang2022catchlive} with what occurred earlier in the broadcast. As audio describing videos typically occurs during post-processing and requires additional editing, existing methods posed by prior work for novice use are difficult to use in real time~\cite{youdescribe,pavel2020rescribe,liu2022crossa11y,wang2021toward,branje2012livedescribe}. Recent work explored sonifying live tennis matches~\cite{jain2023towardslbw}, but domain-specific sonification strategies do not exist for the wide variety of streamed content.
A long history of radio sports broadcasts, in which experienced announcers verbally describe a game, demonstrate that describing real-time descriptions can be understandable and engaging. Livestreams of video game tournaments often feature announcers who verbally describe in-game action.
Building on prior success of describing live events, we investigate the potential for audio description novices who are experts in their domain of interest to produce live descriptions.

\textbf{Synchronous interactions}: Livestreams have remote and synchronous interactions, as opposed to recorded videos that are remote and asynchronous~\cite{johansen1988groupware}. Similar to prior work on watching TV with others (i.e. ``social TV''~\cite{cesar2011past}), community members are able to interact synchronously with each other to build interpersonal relationships~\cite{hamilton2014streaming}. Livestreamers may also interact with their audience by reading chat messages or automated on-screen notifications (\textit{e.g.}, listing a new subscriber) and responding verbally or adapting their actions in response (\textit{e.g.}, ``Thanks for the suggestion, I will try to make the background a farm.''). To encourage interactions, streamers often complement the main streamed content with webcam videos of themselves or their environment, as well as additional on-screen overlays such as subscriber, question, or chat notifications burned into the video feed using OBS Studio~\cite{obs}, Streamlabs~\cite{streamlabs}, or StreamYard~\cite{streamyard}. For streamers with visual impairments, it can be challenging to set up such a streaming environment~\cite{jun2021exploring}. For viewers, it can be difficult to access these elements as they are not screen reader accessible or not directly described by the streamer. 

\textbf{Conversation on and off the streaming platform:} Hamilton et al. described that the use of psuedonyms and text chat can promote self-disclosure that can help people build relationships~\cite{hamilton2014streaming}. People may also carry the same psuedonyms onto shared community spaces outside of streams (\textit{e.g.}, on Discord\footnote{https://discord.com}) to continue to talk to others. We focus our study on content on the streaming platform as the precursor to other types of interactions.

\subsection{Crowdsourcing Accessibility}
Professional audio describers create highly polished audio descriptions for movies that involve scripting, voiceover, and editing to create the finished product. Given a limited amount of expert describers and the high cost of this process, professional description is not practical for user-generated videos. YouDescribe~\cite{youdescribe} offers an approach for people to request descriptions and for volunteer describers to provide descriptions. Prior work has also explored crowdsourcing for answering visual questions~\cite{bigham2010vizwiz}, providing captions and transcriptions (\textit{e.g.}, transcription services like Rev.com), and providing on-demand visual support~\cite{bemyeyes}. However, professionals and crowd workers without domain expertise alike may have difficulty describing content that is unfamiliar to them. For example, Pavel et al.'s formative work with audio describers revealed that describing a new domain can require extensive research into the domain and terminology before providing accurate descriptions~\cite{pavel2020rescribe}.

Instead of crowdsourcing, prior work in community sourcing~\cite{heimerl2012communitysourcing} and learner sourcing~\cite{kim2015learnersourcing} explore drawing from a pool of workers that might have expertise or vested interest in the relevant domain. This approach has had prior success in creating captions. For example, YouTube Community Captions provided community members the chance to add captions to recorded YouTube videos. Prior research also invited domain experts (student learners) who were not experts in providing captions to provide captions that accurately reflected the domain in real-time~\cite{lasecki2012real}. We explore community-sourcing for providing descriptions for livestreams --- a task that requires domain expertise to complete.

\section{Describer Study}

Prior work has explored current challenges and approaches to authoring descriptions for visual media including slide presentations~\cite{peng2021say, peng2021slidecho} and recorded videos~\cite{pavel2020rescribe,natalie2020viscene,wang2021toward}. Livestreams necessitate live description (i.e. written synchronously), rather than asynchronous description of recorded videos. Livestreams often also feature a wide variety of content that requires domain expertise to describe (\textit{e.g.,} complex multiplayer gameplay), a breadth and depth of content that expert describers may not be familiar with. 
To explore the opportunities and challenges of live, community-driven descriptions, we invited 18 livestream viewers with domain expertise to describe livestreams in their domain of interest.

\subsection{Methods}
We conducted a remote within-subjects study with 18 participants describing videos in their domain of expertise across 7 categories. To determine the optimal method of recording live descriptions, we used three description approaches: two synchronous description input methods (one via text and one via speech) and one asynchronous description method (via text). Each participant participated in an individual, 1 hour long, remote study via Zoom (n=4) or Discord (n=14) voice call, and we compensated participants \$20.

\subsubsection{Participants} We recruited 18 sighted participants (P1-P18) from Discord servers and Reddit. All participants were between the ages of 19 and 30 (median=21). Participants ranged from watching 30 minutes to 30 hours of livestreams per week. The participants with the two highest watchtimes per week, 28 and 30 hours, were streamers themselves or frequently watched streams while they performed other tasks. Participants reported their genders as: 11 male, 5 female, and 2 N/A or Non-Conforming. All participants were self-expressed experts in the video category they described and had not previously authored audio or text descriptions for videos. 

\subsubsection{Videos} 
To explore a variety of content, we first selected 7 popular livestream categories from Twitch, a popular livestreaming platform for viewers with visual impairments~\cite{jun2021exploring} and participatory communities ~\cite{hamilton2014streaming}. 
The videos selected spanned video games (League of Legends, Smash Bros, Valorant, The Legend of Zelda: Breath of the Wild (BOTW)), board games (Chess), and creative work (Digital Art, Makeup). As video games represent the most common type of livestream, we selected a variety of video games: a multiplayer online battle arena game (League of Legends), a first-person shooter game (Valorant), a third-person fighting game (Smash Bros), and a single-player adventure game (The Legend of Zelda: Breath of the Wild). For each video category, we selected three livestreams from three different streamers for a total of 21 videos to represent a variety of livestream styles (Table~\ref{tab:livestreams}). We selected a 5 minute clip from each video for the study. 
For five of the livestream categories, we recruited three participants with expertise in the category (Chess, Digital Art, League of Legends, Super Smash Bros., Valorant); for one of the livestream categories we recruited two participants (Makeup); and for one of the livestream categories we recruited one participant (Breath of the Wild).
We downloaded the videos from Twitch for analysis.

\subsubsection{Description Approaches} 
During the study we asked participants to use three description approaches: synchronous text description, asynchronous text description, and synchronous audio description. Our description interfaces built on prior systems for creating audio descriptions that enabled description authors to script and edit descriptions using text~\cite{liu2022crossa11y,natalie2020viscene,natalie2023supporting,yuksel2020human}, and record spoken descriptions using audio~\cite{pavel2020rescribe}. Our live text interface (i.e. \textit{synchronous text description}) let describers write text descriptions. It did not enable describers to record their text descriptions using audio or edit descriptions they had already written as such actions are not possible in real-time. 
To investigate the impact of providing extra time on describer preference and rate, we allowed 2x video time (10 minutes) and enabled text editing along with video navigation in our \textit{asynchronous text description} condition. Finally, we accounted for slower typing speeds by adding \textit{synchronous audio description} to let participants dictate rather than type their descriptions. 

\subsubsection{Describer Extension} 
We implemented our description approaches as a Google Chrome Extension that can be used alongside Twitch to enable real-time description authoring (Figure~\ref{fig:chromeextension}, right).
The extension enables describers to watch the video while writing descriptions. 
Describers designate a new description by pressing `Enter' to start a new line and optionally pressing the backslash key (`\textbackslash') to insert a time code of the segment they are about to describe.
Describers can then write their description.
To review their descriptions, describers click on the time code to jump to the corresponding point in the livestream, then read back their text descriptions while rewatching the video. 

\subsubsection{Procedure} We first asked participants to answer a series of demographic and background questions about their experience watching livestreams and audio describing videos. 
To help participants craft useful descriptions, we shared existing audio description guidelines from YouDescribe~\cite{youdescribefaq} and the Audio Description Project~\cite{adpguidelines}, and showed participants example expert descriptions of the Disney's \textit{The Incredibles (2004)}~\cite{incredibles}. The guidelines gave participants instruction on what to describe (e.g. speakers, lighting, facial expressions, on-screen text) and how to describe it (e.g. use present tense, be objective, avoid technical terminology when possible).  
Participants then installed our Google Chrome extension and completed practice descriptions for one livestream using all three description methods. After the practice session, participants completed the study task, describing three 5-minute video clips within their area of expertise, each using a different description approach, provided in a random order. We counterbalanced conditions so that each video was described using each condition only once across all participants. Post-task we conducted a semi-structured interview to collect participant feedback on their strategies for describing the video, challenges they experienced in describing the video, and preferences among description approaches.

\subsubsection{Analysis}
We recorded the studies using Zoom Cloud Recording for Zoom interviews and OBS Studio~\cite{obs} for Discord interviews, then automatically transcribed the videos using Descript~\cite{descript}. We downloaded text descriptions from our server and segmented them into individual descriptions by new lines. For audio descriptions, we transcribed the description recordings and segmented them into individual descriptions by pauses in speech. We marked the beginning of each description as the time code that the description would appear. We analyzed the interviews using affinity diagramming to group quotes into higher level themes: description strategies (e.g., priorities, challenges, commentating), modality (text, audio), timing (sync, async), and future use (e.g., motivation, scenarios, alternate uses).
We analyzed the descriptions by randomly selecting a subset of 300 descriptions from the whole set of 1183 total descriptions produced by participants, then performing open coding to derive 4 higher level themes and 24 subthemes (Table~\ref{tab:codes}). 

\subsection{Results}
Overall, participants wrote 1183 descriptions over 54 total video description sessions with an average of 21.9 descriptions per video ($\sigma=11.5$ descriptions) and 210.3 words per video ($\sigma=143.7$ words).

\textbf{Livestream description strategies.} 
Over a random subset of 300 descriptions, participants primarily described the main content of the stream (266 descriptions), and occasionally described additional visual content including cameras (34 descriptions) and game-specific actions performed by characters (70 descriptions). To describe the main content of livestreams, participants shared information about the high-level context of the stream (i.e. \textit{state} descriptions) and low-level updates as the stream continued (i.e. \textit{play-by-play} descriptions). State descriptions provided context for understanding play-by-play descriptions, and participants would add a new state description whenever a notable update to the entire stream state occurred. For example, P9 provided a state update for a new League of Legends game starting: \textit{``Doublelift is in champ select. His team bans Yuumi, Poppy, Jax, Taliyah, and Pyke. The enemy team bans Master Yi, Katarina, Akali, Lulu, and Fiddlesticks. Doublelift is support and his ADC is hovering Zeri.''} (V11). Participants provided more play-by-play descriptions (215 descriptions) than state descriptions (56 descriptions).

To fit play-by-play descriptions within limited time, participants often used domain-specific terminology to provide real-time updates (\textit{e.g.}, \textit{``Sage plants spike''} -P16). All participants used domain-specific lingo for at least one description. For example, P15 mentioned they used several shorthand terms that refer to controller inputs including \textit{``dair''} for \textit{``down air''} (a type of attack performed by holding down on the controller's left joystick and pressing the A button while the player's character is not grounded), and \textit{``Up B''} (a type of attack performed by inputting a joystick angle and button combination on the player's controller). While such descriptions helped participants fit additional information about the game, participants expressed concern about the use of technical terminology. For example, P6 questioned if viewers would understand the word \textit{``chibi''} they used to describe a Japanese art style where characters are drawn with exaggerated features.
While participants had domain specific terminology for some in-game actions, participants also mentioned that they occasionally did not know how to describe actions they were seeing (\textit{e.g.}, complex action sequences that used glitches or exploits (P1), character poses (P7), or streamer's facial expressions (P12)), or may not be able to understand complicated action sequences that they were seeing (P4).
On the other hand, participants noted it was easiest to describe objects and actions that were not domain-specific. For example, human body parts in a drawing (P7), common actions like running, swimming and shooting a bow in a game (P1), reading on-screen text verbatim (P4), or describing simple visuals (\textit{e.g.}, a single person on screen). 

Participants identified that low level, play-by-play descriptions were not always the best strategy to describe fast-paced streams or to capture important visual information. Participants responded by changing the level of granularity. For example, the pace of the chess stream on puzzles (V6) was too fast to type or speak each piece movement, so P2 described the stream by mentioning the number of puzzles completed and the number of mistakes the streamer had made. When describing art content, P6 noted that they changed their description strategy from low-level stroke-by-stroke descriptions to higher-level descriptions of what was being drawn: \textit{``Just saying it's being drawn isn't really that helpful. Towards the end, I was trying to say like, the wings are open as if imposing, so that they can sort of imagine it's this big, otherworldly-type figure.''}. Trying to add context for low-level moves in a Valorant game, P18 added commentary that could describe streamer intentions for using certain abilities or aiming certain locations. P14 mentioned that that providing descriptions felt similar to esports commentating. While commentators may provide inspiration for the style and content of the descriptions, P15 higlighted that commentating and describing serve different purposes: \textit{``Commentating is just supplementing what people can see on the screen.''}

While participants all prioritized describing the main content, they included information about other visual streams as possible, when relevant, or in reaction to unidentified sounds. P5 described: \textit{``I'd focus mainly on [...] what they were drawing, then second priority their face cam, and third priority anything else.''}. 15 of 54 sessions started with descriptions of the stream's environment in addition to the main content, but most participants only described parts of the livestream other than the main content when relevant. For example, P12 mentioned that when describing a makeup video, they did not describe the background of the streamer until the streamer directly referenced background objects or walked off-screen. Other participants highlighted that they described on-screen overlays and chat only when mentioned by the streamer or when overlays prompted an unidentified noise. However, when reflecting on their performance, P5 noted that it may have been easier to follow their description if they had described the status of the stream as a whole before starting: \textit{``I would've said, in the top left there's the face cam, below that is the dog face cam, and to the right side of the screen is just the drawing.''} P5 and P11 noted that balancing the streams was difficult due to not knowing what to prioritize (P11) or needing to pay attention to multiple screens (P5). 

\textbf{Comparing livestream description methods.} Overall, participants ranking the description methods from 1 (most preferred) to 3 (least preferred) ranked asynchronous text descriptions as the most preferred input method ($\mu=1.5$, $\sigma=0.62$) followed by synchronous audio ($\mu=2$, $\sigma=0.91$) and synchronous text ($\mu=2.33$, $\sigma=0.69$). 
A Friedman test\footnote{We used Friedman and Wilcoxon tests due to ordinal data (preferences), and non-normal distributions (description count and description words per video minute).} indicated a significant difference in preference between description methods ($\chi^2(2)=6.12$, $p<0.05$), with a post hoc Wilcoxon test with Bonforroni correction indicating a significant difference only between asynchronous and synchronous text descriptions ($p<0.01$). 
Participants also produced more descriptions per video minute with synchronous audio ($\mu=6.28$, $\sigma=2.92$) and asynchronous text ($\mu=4.22$, $\sigma=2.36$) than they could with synchronous text ($\mu=3.20$, $\sigma=1.29$).
Similarly, participants produced more description words per video minute with synchronous audio ($\mu=60.97$, $\sigma=35.52$) and asynchronous text ($\mu=43.70$, $\sigma=24.37$) than they could with synchronous text ($\mu=26.90$, $\sigma=10.18$). 
Friedman tests indicated significant differences in description counts ($\chi^2(2)=18.77$, $p<0.001$) and description words ($\chi^2(2)=17.44$, $p<0.001$) between description methods. Post hoc Wilcoxon tests with Bonforroni correction indicated significant differences ($p<0.05$) between all pairs of methods for both description counts and description words per video minute.

\textit{Text vs. audio descriptions.} 11 participants preferred synchronous audio to synchronous text. 6 participants expressed that speed was the key limitation for text-based methods, and P14 mentioned that their typing was error-prone. To keep up with synchronous text streams, 8 participants reported that they used hotkeys and shorthand. As P5 described, \textit{``If you know their subscriber effects, you can write it once, and then you can just copy-paste it.''}.
5 participants expressed that attempting to avoid talking at the same time as the streamer was the key challenge of dictating audio descriptions. As P12 described: \textit{``The audio was just so difficult. [...] I felt like I was butting into a conversation.''}. On the other hand, when P12 was using text without looking for gaps, \textit{``I felt like I was much more descriptive and tackling more of the things that I'm supposed to be describing rather than just like, this is what's happening.''}. Participants also expressed the challenge of unpredictability of the length of the gap between speech: \textit{`There were moments where I would have a rather long thought about how I would describe [the stream], but I would have to stop because the streamer would start talking''} (P6). P13 noted that they would describe while the streamer focused on the game, but they didn't know when the streamer's focus would break and they start talking again (V17).

\textit{Synchronous vs. asynchronous text descriptions.} 12 participants preferred asynchronous text over synchronous text, and 1 participant rated them equally. Participants preferred asynchronous text as it let them focus on important parts of the stream (P10), pause the video (P5, P6, P7), and not have to describe the video perfectly the first time (P6).
3 participants did not pause more than 3 times during their asynchronous text video, including P2, who preferred \textit{synchronous} text as it felt ``more accurate'' to what they wanted to say. As participants had to budget their own time for asynchronous text, 1 participant ran out of time and only described 3.5 minutes of the 5 minute clip.

8 participants reported that synchronous text descriptions added time pressure to write something down in the moment before there was something else to describe. P16 reported that \textit{``I was gonna type some stuff, but then 40 other things also happened and like we already moved on and I was like, no, I'm just not gonna talk about this anymore.''}. As P15 described: \textit{``I almost feel bad. I feel like there were details that would be nice to know that I just wasn't able to say.''}

\textbf{Future description.} Participants reported that composing descriptions was challenging and that they would be willing to describe videos again in the future depending on how interested they are in the video. While most participants preferred to describe videos they would watch anyway, P18 reported that they would prefer to describe videos they are \textit{not} as interested in so that they can focus on enjoying their streams of interest. 7 participants reported that they would provide descriptions if compensated (\textit{e.g.} by the streamer), while 11 participants would volunteer to write descriptions. P3 compared writing descriptions to chat moderation, which is often a volunteer task. As describing is challenging, several participants mentioned that they would want to describe in smaller blocks of time, from around 15 minutes (P1) up to an hour at once (P5, P7, P16) for synchronous text.
5 participants suggested alternate use cases for using written descriptions as sighted people, including watching a stream in the background or on another monitor (3 participants), walking outside without their phone out (1 participant), driving (1 participant), or getting ready for an event (1 participant).

\section{Audience Study}
We conducted a study with livestream viewers with visual impairments to learn about current livestream viewing practices and challenges and surface description preferences.

\subsection{Methods}
We conducted a 1 hour remote study via Zoom with 9 participants with visual impairments who used screen readers to access their device. 
Participants were recruited through Reddit discussion boards~\cite{blindsurveys} and email lists, and all participants had used Zoom in the past. 

We compensated participants \$25 for their time.

\subsubsection{Participants}
Participants U1 through U9 ranged from ages 27 to 57 (6 male and 3 female) (Table~\ref{tab:participantsblv}). All participants reported that YouTube was their primary video streaming platform, with one participant watching Twitch an equal amount. Participants spent on average of 0.25 hours to 10 hours per week watching live video. 

\subsubsection{Procedure}
\begin{figure}
    \centering
    \includegraphics[width=3.33in]{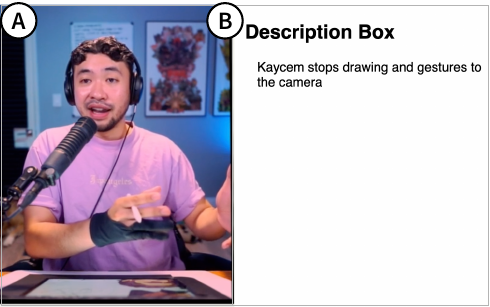}
    \caption{Description playback application featuring a section of the source video (A) and the description box (B). The description box updates as the playback time of the video matches a timecode for a description. If the text within the description box is updated, a sound effect plays and the new description is automatically read out loud by the participant's screen reader. Source video: \textit{how to IMPROVE your SKILLS QUICKLY + NEW SUB GOAL?!? !bootcamp !youtube} by Kaycem~\protect{\cite{kaycem}}}
    \Description[A figure of the description playback application]{Figure 2: A figure depicting the prototype application for playing back visual descriptions of Twitch streams. On the left side of the figure is a still screenshot of a Twitch stream marked with the letter A. On the left side of the stream is a man facing the camera wearing headphones, holding an Apple Pencil and gesturing towards the camera. He speaks into a microphone on a boom arm in front of him. On the desk lying flat in front of him is an iPad. On the right side of the screen is a box titled ``Description Box'' marked with the letter B, which includes the text below it: ``Kaycem stops drawing and gestures to the camera.''}
    \label{prototype}
\end{figure}

We first asked participants demographic questions and background questions about their current livestream watching practices, platform and content accessibility challenges, and strategies for gaining more information.
To demonstrate current practice, participants then searched for, selected, and watched 5 minutes of any one livestream on their preferred livestream viewing platform. We invited participants to ask questions about the visual content in the video and to rate their perceived accessibility of the video from 1 (very inaccessible) to 7 (very accessible), similar to Liu et al.~\cite{liu2021what}.
To provide feedback on sample descriptions, participants selected one topic from the 7 livestream categories in Section 3.1.2 and watched 3 different five-minute clips on that topic. We paired each of these 3 streams with a description from a different description approach (synchronous text, synchronous audio, asynchronous text) produced during the Describer Study. All participants selecting the same category were served the same video-description approach pairs in a random order. 
Participants accessed descriptions via links to a webpage displaying a recording of the video and a description box~(Figure \ref{prototype}). Before each video, the researcher provided an overview on the video context, including a brief description of the streamer and the main content of the clip. 
Once participants began watching each clip, descriptions were read back automatically by the participant's screen reader as the corresponding timestamp in the video overlapped with a stored description's time code. To control for audio quality and noise, all descriptions created with the synchronous audio description method were transcribed and played back as if they were written via text.
After each stream, we invited participants to ask questions about the visual content in the scene, rate their perceived accessibility of the video with and without descriptions from 1 (very inaccessible) to 7 (very accessible), and provide feedback on what they liked, disliked, and wished to improve about the descriptions.
Finally, we asked participants closing questions about their overall livestream description comparisons and preferences.


\subsubsection{Analysis}
We asked participants to screen share with sound using Zoom, recorded the studies using Zoom Cloud Recording, then automatically transcribed the videos using Microsoft Office Word 365~\cite{word365} and Adobe Premiere Pro CC~\cite{premiere}. We grouped participant responses according to our questions (e.g., current practice, strategies, challenges, and description preferences), then iteratively identified concepts using open coding.

\subsection{Results}
Overall, participants rated the accessibility of their preferred streaming platform as 4.7 ($\sigma=1.2$) out of 7 and similarly rated the live content on these platforms as 4.2 ($\sigma=1.5$) out of 7. Participants rated the videos they hand-selected during the co-watching study from streamers they were familiar with as 5.78 ($\sigma=1.39$). For our example descriptions to probe for feedback, 5 participants chose to watch Breath of the Wild (BOTW), 2 participants chose digital art, 1 participant chose chess, and 1 participant chose Super Smash Bros. Participants rated the videos they watched as 2.3 ($\sigma=1.2$) without descriptions and 5.2 ($\sigma=1.4$) with descriptions.

\subsubsection{Current livestream viewing practices}
Participants primarily watched livestreams to gain information (U1, U2, U3, U4, U5, U6, U8, U9) or for entertainment (U1, U2, U3, U4, U6, U7, U8). Live videos for gaining information included live news (U6, U9), travel (U1, U3), online conferencing (U5), learning guitar (U3), cooking (U8), household repair (U8), learning game strategy (U9), and other personal interests and hobbies (U2, U3, U4, U5). Domains for live videos in entertainment included gaming (U1, U4, U6, U7, U9), music (U1, U2, U3, U8), podcasts (U2, U6), Q\&A's (U1, U3), and general commentary (U3, U7). 
Participants reported that they watched livestreams in particular as they appreciated the ability to interact in real-time with the presenter and other viewers, including the ability to ask questions and receive information at the same time as recorded and alongside everyone else (U5, U6, U7, U8, U9). Participants reported that livestreams included ``more honest reactions'' (U1) from streamers compared to typical edited content (U4).
Participants also appreciated learning more about other people's experiences (U1, U2), including culture (U1), travel (U1, U2), or catching up with their friends (U2). 

To find livestreams to watch, only U3 and U4 reported using recommendation feeds to find live videos of interest, unlike prior work for recorded videos in which most people used their recommendations~\cite{liu2021what}. Instead, participants watched streams shared by their friends (U5, U7), users on other social media (U2, U7), or news sites (U9). Many participants also monitored notifications from channels they follow, tuning in when they go live (U1, U2, U4, U6, U9). Otherwise, participants would search for their hobbies or specific topics they're interested in and pick one of the top results (U2, U5, U6, U8). During the co-watching portion, 5 participants used YouTube search to look for specific topics or streamers they regularly watch; U4 selected a recommendation from a subscribed channel on the front page of YouTube; U5 used a recommendation from an email mailing list; U7 checked their Twitch following list, but no one was online, so they used Twitch search for specific streamers they are familiar with; and U8 used Google search and appended ``YouTube'' to their query.

\subsubsection{Current accessibility of livestream content \& platforms.} Participants reported that livestreams were most accessible when they had good audio quality, clear voices, a lack of background music (U5, U6), extensive narration from the streamer (U3, U8), and lack of fast-paced action (U3, U5, U8).
U6 and U9 both picked accessible audio game streams with no visuals and presented by a visually impaired streamer--- these streams were completely accessible to them. 
Some reasons that livestreams were inaccessible were similar to prior work exploring the accessibility of recorded video~\cite{liu2021what}, including: on-screen text burned into the video but not described, unclear visual references (\textit{e.g.}, ``this'', ``there''), unidentified sounds, and lack of description of the main visual content. However, livestreams posed additional challenges: First, unexplained sounds were frequent due to sound-producing overlays added to the video (U5, U6) (\textit{e.g.}, a subscriber notification). Additionally, as livestreams are long and unedited compared to recorded videos, streamers often left long silences as they took a break from talking (U3, U4, U5, U6)--- or they would break from talking about the game to talk about miscellaneous topics, such as telling a story or responding to chat, that could make it difficult to follow the main content (U2, U3, U7). While watching a stream, U2 commented: \textit{``I'm not sure if he's showing anything or if he's just talking or I have no idea here.''}
Most participants noted that chat messages were particularly hard to read due to factors like the speed of the chat and custom emotes (i.e. emoji-like images specific to the stream), such that when streamers responded to chat without describing it (\textit{e.g.}, \textit{``Yeah, I agree with that, let's try it.''}), participants were unable to understand the context for the response.
Participants reported that they also wanted more information about the streamer, including facial expressions and body language (U3), as well as what they look like (U8). U7 mentioned that when streamers were playing games, they wanted more background information about the game status (\textit{e.g.}, the place on the map, the damage updates) that were typically not included in streamer narrations: \textit{``I can hear that they're taking damage, but without a low health indicator noise you don't know how low they are.''} U9 noted that when it was a game that they were not familiar with, it was difficult to learn what was going on.

Finally, the livestream platforms themselves were not fully accessible. 4 participants who used YouTube to watch livestreams mentioned wanting to watch Twitch streams outside of the study, but found the platform difficult to use due to poor labeling of interface elements and difficulty navigating using a screen reader.

\subsubsection{Strategies for gaining information about inaccessible streams.} 
Participants mentioned strategies for handling inaccessible streams including: moving on to find another stream, asking the streamer or audience for additional information (on Discord or chat), prompting the streamer to change their narration style, and asking friends or family. When participants were not invested in a particular stream, participants indicated they would move on to find other streams (U2, U5, U6, U8): \textit{``It's simple. I don't watch it. I mean, if it's gonna frustrate me, so some people might get mad and rant about it. I'm like, OK, I can't watch it''} (U2). Participants U2, U4, U5, U6, and U9 reported reaching out to the streamer directly in chat or via email to provide more complete narrations for their actions. U4 mentioned that \textit{``I have been known to reach out to the video provider to the video upload and say `hey, I'm a blind individual consuming your content. Tell me what you're doing. Tell me what you're seeing.''}. U4 reported that most of the streamers they follow on Twitch are good about trying to cue their viewers into what they're doing, and that U4 would remind them when the streamer forgets. Participants also suggested looking in chat or asking other viewers questions via chat (U1, U3, U5, U6, U7), though the chat itself was difficult to navigate. U7 described gaining additional information via live chat: \textit{``One time I was on LilyPichu’s stream and I asked why there's a tomato emoji after the name of the stream that day. And I was like, I'm blind and I'm just curious. And a few people said it's because she dyed her hair red and oh, okay. But it was hard to find that in the massive stream of faces with tears of joy.''} U7 also mentioned they look at the chat when joining a stream as people often comment on what is going on in the stream, but neither U7 (nor U6) send a message themselves unless they are particularly curious about something, as they see it as bothersome.
Participants also asked sighted friends and family members to answer visual questions (U3, U6). Finally, participants used external online sources such as web search (U5, U7, U9), Twitter (U5), and wikis (U7) to learn more about the context for the stream.

\subsubsection{Livestream description preferences.} Participants all reported that they wanted descriptions to prioritize describing the main content. Additional descriptions of visual content should be described as relevant, including: reading out pop up overlays with text (U1), the appearance of the streamer (U2), and streamer reactions from facial expressions (U5). This preference order reflected community members' stated priorities in creating their descriptions.

\textit{Description content.} From community member-provided descriptions, all participants reported that the descriptions provided useful information for understanding the stream. As a result, most participants reported that the accessibility of the video improved after descriptions. U1 summarized: \textit{``Most of the time you're watching the video when you're [...] totally blind and you have no idea what's going on; having any form of description is helpful.''} U2 highlighted that such descriptions were particularly useful for inaccessible moments: \textit{``For streams where the streamer is talking aimlessly or about something independent of the main activity they're performing, it makes a stream someone would otherwise click off to something interesting.''}. Participants mentioned that they liked when the describer included information about the player's strategy instead of only raw visual content--- contrary to audio description guidelines. Participants also found the contextual information (\textit{e.g.}, global descriptions of the video context before the start of the video) to be helpful in understanding the content of the video. Participants offered additional information they may want to know about for state descriptions including how fast the chat is moving (U8) and the goals of the gameplay (U5, U7).
At times, participants disagreed with a describer's choice to include additional information (e.g., describing a subscriber notification in the middle of a drawing stream). Other participants reported that they wanted additional information about background about the game (U5) and additional detail about the minutiae of the game (U3, U8, U6). 

\textit{Expertise of describers and terminology.} Most participants reported that they wanted descriptions from people with familiarity with the visual content as \textit{``otherwise you can't learn from the stream''} (U2). U1 reported that they had heard descriptions with errors from people who lack experience in the past, and U7 mentioned that they wanted describers to be fans of the media: \textit{``The more that the person's passionate about it, the better they do tend to be.''} Participants also clarified that while many domains require expertise (\textit{e.g.}, chess, art), expertise may not be necessary for some types of content (\textit{e.g.}, a travel video). 

U5 said that when audience members are coming to the video \textit{``with zero background knowledge''}, the descriptions should be as \textit{``beginner mode''} as possible; however, the describers in our study frequently used domain-specific terminology. While most participants found the terminology to be understandable from prior knowledge or context cues, other participants reported that they would like the ability to gain more information about terms used (\textit{e.g.}, via a wiki link -U5). To make the video understandable to people with different skill levels, U9 suggested that descriptions contain no more than around 30\% terminology so that there are enough context cues to understand the terms used.

\textit{Description format.} Participants appreciated the text format of the descriptions as they could use their existing screen reader settings and reread descriptions when they were confused. However, participants said that they would like the ability to easily refer to past descriptions and a mode where they can pause the video to read descriptions in isolation. At times, text to speech pronounced words incorrectly, especially if the word was domain-specific or contained a spelling error brought on by the describer. U1 clarified that it would take them extra time to understand mispronunciations, but that they ultimately understood the content. 

As we presented both forms of description provided by community members (text and speech) as screen reader-accessible text for consistency (\textit{e.g.}, screen reader speed, audio quality, and background noise), the most critical description feedback was for the voice-generated descriptions. In particular, participants found these descriptions to contain awkward pauses, long gaps in description, incomplete sentences, and filler words (um, uh) that made them more difficult to understand.

\textit{Description timing.} Participants all wanted the ability to be able to keep up with the stream in real time. Participants reported that they would tolerate delays from ``within 15 seconds'' (U4) to up to ``1-2 minutes'' (U7). U5 mentioned that they wanted the option to manually adjust the video delay to catch up with descriptions.
U3 summarized that synchronized descriptions had the effect of: \textit{``I'm right here. I'm participating in the moment. This is what's happening. [...]
it's really cool to feel like I'm an equal participant in that moment'' (U3)}. U1 mentioned \textit{``I think I'm just looking forward to something like this becoming mainstream. With more accessibility and primarily more description for videos either in a text format or human narrated description.''}.

\section{Discussion}
Livestream viewers with visual impairments reported that they strategically used social support from the livestream community and their network to make livestreams accessible, but that they valued additional information provided by community member descriptions. Our studies point to additional ways to make livestreams more accessible to viewers with visual impairments. 

\subsection{Improving Live Descriptions In Practice} Livestream viewers found community member descriptions useful for understanding the visual content in a stream. Our studies indicate additional ways to help community members craft high-quality descriptions in practice:

\subsubsection{Synchronous Listener-Describer Communication} Our study imitated live description and viewing experiences (\textit{e.g.}, inability to skip ahead, timing), and in future work we will explore fully synchronous description. Audience members' current synchronous use of chat and Discord to gather information and provide feedback suggests that future two-way listener-describer communication may be beneficial to improve descriptions. For example, listeners could discuss their expertise and preferences with describers apriori, provide live description feedback, or ask live visual questions.

\subsubsection{Maintaining Description Consistency} Community members reported they could describe 15 minutes to 1 hour at a time, such that multiple describers will be required to produce descriptions for streams that are many hours long. On the other hand, our audience study indicated that \textit{description consistency} is important, as new terminology takes time to learn. 
To maintain description consistency between describers, future work will explore creating per-domain hotkey libraries of common characters and actions (similar to copy-paste strategies used by describers in our study) or providing describers a warm-up period where they observe the prior describers' descriptions.

\subsubsection{Sticky State Descriptions}
Community members used \textit{state} descriptions to set the scene of the stream and subsequent \textit{play-by-play} descriptions to deliver updates. However, \textit{play-by-play} descriptions were often not understandable without corresponding state descriptions, such that viewers joining mid-stream may be confused. Future automated or human-authored descriptions should include ``sticky'' state descriptions that persist until new state descriptions are authored.
When livestream viewers join a stream, they could first listen to the sticky state description before listening to play-by-play updates. Providing state descriptions alone, rather than full live descriptions, may also provide an easier task for first-time describers. 

\subsubsection{Tailoring Descriptions} Livestream viewers reported a variety of preferences in the level of detail they wanted from descriptions and the level of expertise they wanted the describer to have. In the future, we will explore automatically classifying descriptions (\textit{e.g.}, based on our codes) and either provide viewers the ability to selectively toggle certain types of descriptions (\textit{e.g.}, streamer appearance, state vs. play-by-play descriptions), or learn from viewer preferences indicated via lightweight feedback (\textit{e.g.}, skipping the description, or thumbs up/thumbs down). 

\subsubsection{Transcription Accuracy}
While several describers preferred audio input, livestream viewers noted that descriptions transcribed from audio were less clear than descriptions written via text. To improve the delivery of descriptions produced via live audio description, future work will explore automatically removing filler words and repetitions from audio input using a large language model. Future work will also explore improving text transcription errors for uncommon words by using a library of common domain actions and characters to inform transcription. 

\subsection{Beyond Community Descriptions} 
While livestream viewers found descriptions useful, they suggested additional approaches to make livestreams more accessible, including filtering chat messages and prompting streamers to describe visual content ahead of time. 
To highlight useful chat messages, future work may surface visual information in the livestream chat to augment descriptions~\cite{huh2022cocomix} or identify important chat messages~\cite{yang2022catchlive}.
To help streamers remember to describe their stream, future work may consider automated approaches to prompt streamers to describe missing visual information~\cite{peng2021say,liu2022crossa11y}.
In the future, we will also explore how to augment community descriptions with multiple modalities (\textit{e.g.} sonification~\cite{jain2023towardslbw,holloway2022infosonics,kramer2010sonification,walker2011theory}, haptics~\cite{yuan2008blind,mcdaniel2013evaluation}), and help livestream viewers surface accessible livestreams~\cite{liu2021what}. For example, we will explore modifying Liu et al.'s~\cite{liu2021what} approach for surfacing accessible YouTube videos to provide the accessibility score of a stream in progress, or for the streamer's prior streams, before a viewer decides to join.

\subsection{Study Limitations} While our descriptions are usable by all screen reader users, livestream viewers in our study were all blind (n=5) or blind with light perception (n=4). Future work should investigate livestream viewing practices and challenges of low vision viewers, and explore approaches to provide support for low vision viewers. 
For example, community members could link their descriptions to a spatial region such that low vision viewers may play back descriptions relevant to their current zoomed-in view.
Our studies featured video categories related to the expertise of sighted describers. While such descriptions produced useful description feedback in the audience study, our future work will explore studying synchronous descriptions with matched interests. 
Our studies explored the feasibility of community-driven live descriptions in the context of short, 1 hour study sessions. Our work reveals an opportunity for longer-term studies to study the impact of fatigue and expertise gain while using and providing descriptions over longer periods of time.

\section{Conclusion}
We conducted two studies exploring community-driven descriptions to improve the accessibility of livestreams. Our work reveals that livestream community members with visual impairments already use social support from the livestream community to gain additional information. However, community-driven descriptions would provide a valuable, additional channel for livestream viewers to gain access to livestream information, as well as the community building afforded by live participation. As our work is the first to investigate describing livestreams in real-time, we identify areas for future work in improving community-driven descriptions and general livestream accessibility. As we communicate via new mediums such as livestreams, we must assure that they are accessible to everyone. We hope that our work will catalyze future research and systems to improve livestream accessibility.

\bibliographystyle{ACM-Reference-Format}
\bibliography{biblio}


\begin{thebibliography}{54}


\ifx \showCODEN    \undefined \def \showCODEN     #1{\unskip}     \fi
\ifx \showDOI      \undefined \def \showDOI       #1{#1}\fi
\ifx \showISBNx    \undefined \def \showISBNx     #1{\unskip}     \fi
\ifx \showISBNxiii \undefined \def \showISBNxiii  #1{\unskip}     \fi
\ifx \showISSN     \undefined \def \showISSN      #1{\unskip}     \fi
\ifx \showLCCN     \undefined \def \showLCCN      #1{\unskip}     \fi
\ifx \shownote     \undefined \def \shownote      #1{#1}          \fi
\ifx \showarticletitle \undefined \def \showarticletitle #1{#1}   \fi
\ifx \showURL      \undefined \def \showURL       {\relax}        \fi
\providecommand\bibfield[2]{#2}
\providecommand\bibinfo[2]{#2}
\providecommand\natexlab[1]{#1}
\providecommand\showeprint[2][]{arXiv:#2}

\bibitem[Adobe(2022)]%
        {premiere}
\bibfield{author}{\bibinfo{person}{Adobe}.} \bibinfo{year}{2022 (accessed Dec 13, 2022)}\natexlab{}.
\newblock \bibinfo{title}{Premiere Pro}.
\newblock
\newblock
\urldef\tempurl%
\url{https://www.adobe.com/products/premiere.html}
\showURL{%
\tempurl}


\bibitem[Bigham et~al\mbox{.}(2010)]%
        {bigham2010vizwiz}
\bibfield{author}{\bibinfo{person}{Jeffrey~P Bigham}, \bibinfo{person}{Chandrika Jayant}, \bibinfo{person}{Hanjie Ji}, \bibinfo{person}{Greg Little}, \bibinfo{person}{Andrew Miller}, \bibinfo{person}{Robert~C Miller}, \bibinfo{person}{Robin Miller}, \bibinfo{person}{Aubrey Tatarowicz}, \bibinfo{person}{Brandyn White}, \bibinfo{person}{Samual White}, {et~al\mbox{.}}} \bibinfo{year}{2010}\natexlab{}.
\newblock \showarticletitle{Vizwiz: nearly real-time answers to visual questions}. In \bibinfo{booktitle}{\emph{Proceedings of the 23nd annual ACM symposium on User interface software and technology}}. \bibinfo{pages}{333--342}.
\newblock


\bibitem[Branje and Fels(2012)]%
        {branje2012livedescribe}
\bibfield{author}{\bibinfo{person}{Carmen~J Branje} {and} \bibinfo{person}{Deborah~I Fels}.} \bibinfo{year}{2012}\natexlab{}.
\newblock \showarticletitle{Livedescribe: can amateur describers create high-quality audio description?}
\newblock \bibinfo{journal}{\emph{Journal of Visual Impairment \& Blindness}} \bibinfo{volume}{106}, \bibinfo{number}{3} (\bibinfo{year}{2012}), \bibinfo{pages}{154--165}.
\newblock


\bibitem[Cesar and Geerts(2011)]%
        {cesar2011past}
\bibfield{author}{\bibinfo{person}{Pablo Cesar} {and} \bibinfo{person}{David Geerts}.} \bibinfo{year}{2011}\natexlab{}.
\newblock \showarticletitle{Past, present, and future of social TV: A categorization}. In \bibinfo{booktitle}{\emph{2011 IEEE consumer communications and networking conference (CCNC)}}. IEEE, \bibinfo{pages}{347--351}.
\newblock


\bibitem[Chen et~al\mbox{.}(2021)]%
        {chen2021afraid}
\bibfield{author}{\bibinfo{person}{Xinyue Chen}, \bibinfo{person}{Si Chen}, \bibinfo{person}{Xu Wang}, {and} \bibinfo{person}{Yun Huang}.} \bibinfo{year}{2021}\natexlab{}.
\newblock \showarticletitle{"I was afraid, but now I enjoy being a streamer!" Understanding the Challenges and Prospects of Using Live Streaming for Online Education}.
\newblock \bibinfo{journal}{\emph{Proceedings of the ACM on Human-Computer Interaction}} \bibinfo{volume}{4}, \bibinfo{number}{CSCW3} (\bibinfo{year}{2021}), \bibinfo{pages}{1--32}.
\newblock


\bibitem[Corp(2023)]%
        {airaio}
\bibfield{author}{\bibinfo{person}{Aira~Tech Corp}.} \bibinfo{year}{2023 (accessed May 2023)}\natexlab{}.
\newblock \bibinfo{title}{Aira}.
\newblock
\newblock
\newblock
\shownote{\url{https://aira.io}}.


\bibitem[Descript(2022)]%
        {descript}
\bibfield{author}{\bibinfo{person}{Descript}.} \bibinfo{year}{2022 (accessed Sep 6, 2022)}\natexlab{}.
\newblock \bibinfo{title}{Descript}.
\newblock
\newblock
\urldef\tempurl%
\url{https://www.descript.com/}
\showURL{%
\tempurl}


\bibitem[Eyes(2023)]%
        {bemyeyes}
\bibfield{author}{\bibinfo{person}{Be~My Eyes}.} \bibinfo{year}{2023 (accessed May 2023)}\natexlab{}.
\newblock \bibinfo{title}{Be My Eyes}.
\newblock
\newblock
\newblock
\shownote{\url{https://www.bemyeyes.com}}.


\bibitem[Faas et~al\mbox{.}(2018)]%
        {faas2018watch}
\bibfield{author}{\bibinfo{person}{Travis Faas}, \bibinfo{person}{Lynn Dombrowski}, \bibinfo{person}{Alyson Young}, {and} \bibinfo{person}{Andrew~D Miller}.} \bibinfo{year}{2018}\natexlab{}.
\newblock \showarticletitle{Watch me code: Programming mentorship communities on twitch. tv}.
\newblock \bibinfo{journal}{\emph{Proceedings of the ACM on Human-Computer Interaction}} \bibinfo{volume}{2}, \bibinfo{number}{CSCW} (\bibinfo{year}{2018}), \bibinfo{pages}{1--18}.
\newblock


\bibitem[Fraser et~al\mbox{.}(2019)]%
        {fraser2019sharing}
\bibfield{author}{\bibinfo{person}{C~Ailie Fraser}, \bibinfo{person}{Joy~O Kim}, \bibinfo{person}{Alison Thornsberry}, \bibinfo{person}{Scott Klemmer}, {and} \bibinfo{person}{Mira Dontcheva}.} \bibinfo{year}{2019}\natexlab{}.
\newblock \showarticletitle{Sharing the studio: How creative livestreaming can inspire, educate, and engage}.
\newblock In \bibinfo{booktitle}{\emph{Proceedings of the 2019 on Creativity and Cognition}}. \bibinfo{pages}{144--155}.
\newblock


\bibitem[Gleason et~al\mbox{.}(2020)]%
        {gleason2020making}
\bibfield{author}{\bibinfo{person}{Cole Gleason}, \bibinfo{person}{Amy Pavel}, \bibinfo{person}{Himalini Gururaj}, \bibinfo{person}{Kris Kitani}, {and} \bibinfo{person}{Jeffrey Bigham}.} \bibinfo{year}{2020}\natexlab{}.
\newblock \showarticletitle{Making GIFs Accessible}. In \bibinfo{booktitle}{\emph{Proceedings of the 22nd International ACM SIGACCESS Conference on Computers and Accessibility}}. \bibinfo{pages}{1--10}.
\newblock


\bibitem[Hamilton et~al\mbox{.}(2014)]%
        {hamilton2014streaming}
\bibfield{author}{\bibinfo{person}{William~A Hamilton}, \bibinfo{person}{Oliver Garretson}, {and} \bibinfo{person}{Andruid Kerne}.} \bibinfo{year}{2014}\natexlab{}.
\newblock \showarticletitle{Streaming on twitch: fostering participatory communities of play within live mixed media}. In \bibinfo{booktitle}{\emph{Proceedings of the SIGCHI conference on human factors in computing systems}}. \bibinfo{pages}{1315--1324}.
\newblock


\bibitem[Heimerl et~al\mbox{.}(2012)]%
        {heimerl2012communitysourcing}
\bibfield{author}{\bibinfo{person}{Kurtis Heimerl}, \bibinfo{person}{Brian Gawalt}, \bibinfo{person}{Kuang Chen}, \bibinfo{person}{Tapan Parikh}, {and} \bibinfo{person}{Bj{\"o}rn Hartmann}.} \bibinfo{year}{2012}\natexlab{}.
\newblock \showarticletitle{CommunitySourcing: engaging local crowds to perform expert work via physical kiosks}. In \bibinfo{booktitle}{\emph{Proceedings of the SIGCHI conference on human factors in computing systems}}. \bibinfo{pages}{1539--1548}.
\newblock


\bibitem[Holloway et~al\mbox{.}(2022)]%
        {holloway2022infosonics}
\bibfield{author}{\bibinfo{person}{Leona~M Holloway}, \bibinfo{person}{Cagatay Goncu}, \bibinfo{person}{Alon Ilsar}, \bibinfo{person}{Matthew Butler}, {and} \bibinfo{person}{Kim Marriott}.} \bibinfo{year}{2022}\natexlab{}.
\newblock \showarticletitle{Infosonics: Accessible infographics for people who are blind using sonification and voice}. In \bibinfo{booktitle}{\emph{Proceedings of the 2022 CHI Conference on Human Factors in Computing Systems}}. \bibinfo{pages}{1--13}.
\newblock


\bibitem[Hopin(2023)]%
        {streamyard}
\bibfield{author}{\bibinfo{person}{Hopin}.} \bibinfo{year}{2023 (accessed May 2023)}\natexlab{}.
\newblock \bibinfo{title}{StreamYard}.
\newblock
\newblock
\newblock
\shownote{\url{https://streamyard.com}}.


\bibitem[Huang et~al\mbox{.}(2017)]%
        {huang2017leveraging}
\bibfield{author}{\bibinfo{person}{Yun Huang}, \bibinfo{person}{Yifeng Huang}, \bibinfo{person}{Na Xue}, {and} \bibinfo{person}{Jeffrey~P Bigham}.} \bibinfo{year}{2017}\natexlab{}.
\newblock \showarticletitle{Leveraging complementary contributions of different workers for efficient crowdsourcing of video captions}. In \bibinfo{booktitle}{\emph{Proceedings of the 2017 chi conference on human factors in computing systems}}. \bibinfo{pages}{4617--4626}.
\newblock


\bibitem[Huh et~al\mbox{.}(2022)]%
        {huh2022cocomix}
\bibfield{author}{\bibinfo{person}{Mina Huh}, \bibinfo{person}{YunJung Lee}, \bibinfo{person}{Dasom Choi}, \bibinfo{person}{Haesoo Kim}, \bibinfo{person}{Uran Oh}, {and} \bibinfo{person}{Juho Kim}.} \bibinfo{year}{2022}\natexlab{}.
\newblock \showarticletitle{Cocomix: Utilizing Comments to Improve Non-Visual Webtoon Accessibility}. In \bibinfo{booktitle}{\emph{Proceedings of the 2022 CHI Conference on Human Factors in Computing Systems}}. \bibinfo{pages}{1--18}.
\newblock


\bibitem[Institute(2019a)]%
        {youdescribefaq}
\bibfield{author}{\bibinfo{person}{The Smith-Kettlewell Eye~Research Institute}.} \bibinfo{year}{2019}\natexlab{a}.
\newblock \bibinfo{title}{YouDescribe FAQ for describers}.
\newblock
\newblock
\newblock
\shownote{\url{https://youdescribe.org/support/describers}}.


\bibitem[Institute(2019b)]%
        {youdescribe}
\bibfield{author}{\bibinfo{person}{The Smith-Kettlewell Eye~Research Institute}.} \bibinfo{year}{2019}\natexlab{b}.
\newblock \bibinfo{title}{YouDescribe.com}.
\newblock
\newblock
\newblock
\shownote{\url{https://youdescribe.org/}}.


\bibitem[Jain et~al\mbox{.}(2023)]%
        {jain2023towardslbw}
\bibfield{author}{\bibinfo{person}{Gaurav Jain}, \bibinfo{person}{Basel Hindi}, \bibinfo{person}{Connor Courtien}, \bibinfo{person}{Xin Yi~Therese Xu}, \bibinfo{person}{Conrad Wyrick}, \bibinfo{person}{Michael Malcolm}, {and} \bibinfo{person}{Brian~A. Smith}.} \bibinfo{year}{2023}\natexlab{}.
\newblock \showarticletitle{Towards Accessible Sports Broadcasts for Blind and Low-Vision Viewers}. In \bibinfo{booktitle}{\emph{Proceedings of the 2023 CHI Conference on Human Factors in Computing Systems}}. \bibinfo{pages}{1--7}.
\newblock


\bibitem[Johansen(1988)]%
        {johansen1988groupware}
\bibfield{author}{\bibinfo{person}{Robert Johansen}.} \bibinfo{year}{1988}\natexlab{}.
\newblock \bibinfo{booktitle}{\emph{Groupware: Computer support for business teams}}.
\newblock \bibinfo{publisher}{The Free Press}.
\newblock


\bibitem[Jun et~al\mbox{.}(2021)]%
        {jun2021exploring}
\bibfield{author}{\bibinfo{person}{Joonyoung Jun}, \bibinfo{person}{Woosuk Seo}, \bibinfo{person}{Jihyeon Park}, \bibinfo{person}{Subin Park}, {and} \bibinfo{person}{Hyunggu Jung}.} \bibinfo{year}{2021}\natexlab{}.
\newblock \showarticletitle{Exploring the experiences of streamers with visual impairments}.
\newblock \bibinfo{journal}{\emph{Proceedings of the ACM on Human-Computer Interaction}} \bibinfo{volume}{5}, \bibinfo{number}{CSCW2} (\bibinfo{year}{2021}), \bibinfo{pages}{1--23}.
\newblock


\bibitem[Kaycem(2023)]%
        {kaycem}
\bibfield{author}{\bibinfo{person}{Kaycem}.} \bibinfo{year}{2023 (accessed July 2023)}\natexlab{}.
\newblock \bibinfo{title}{how to IMPROVE your SKILLS QUICKLY + NEW SUB GOAL?!? !bootcamp !youtube}.
\newblock
\newblock
\newblock
\shownote{\url{https://www.twitch.tv/videos/1854614493}}.


\bibitem[Kim et~al\mbox{.}(2015)]%
        {kim2015learnersourcing}
\bibfield{author}{\bibinfo{person}{Juho Kim} {et~al\mbox{.}}} \bibinfo{year}{2015}\natexlab{}.
\newblock \emph{\bibinfo{title}{Learnersourcing: improving learning with collective learner activity}}.
\newblock \bibinfo{thesistype}{Ph.\,D. Dissertation}. \bibinfo{school}{Massachusetts Institute of Technology}.
\newblock


\bibitem[Kramer et~al\mbox{.}(2010)]%
        {kramer2010sonification}
\bibfield{author}{\bibinfo{person}{Gregory Kramer}, \bibinfo{person}{Bruce Walker}, \bibinfo{person}{Terri Bonebright}, \bibinfo{person}{Perry Cook}, \bibinfo{person}{John~H Flowers}, \bibinfo{person}{Nadine Miner}, {and} \bibinfo{person}{John Neuhoff}.} \bibinfo{year}{2010}\natexlab{}.
\newblock \showarticletitle{Sonification report: Status of the field and research agenda}.
\newblock  (\bibinfo{year}{2010}).
\newblock


\bibitem[Lasecki et~al\mbox{.}(2012)]%
        {lasecki2012real}
\bibfield{author}{\bibinfo{person}{Walter Lasecki}, \bibinfo{person}{Christopher Miller}, \bibinfo{person}{Adam Sadilek}, \bibinfo{person}{Andrew Abumoussa}, \bibinfo{person}{Donato Borrello}, \bibinfo{person}{Raja Kushalnagar}, {and} \bibinfo{person}{Jeffrey Bigham}.} \bibinfo{year}{2012}\natexlab{}.
\newblock \showarticletitle{Real-time captioning by groups of non-experts}. In \bibinfo{booktitle}{\emph{Proceedings of the 25th annual ACM symposium on User interface software and technology}}. \bibinfo{pages}{23--34}.
\newblock


\bibitem[Lee(2011)]%
        {lee2011participatory}
\bibfield{author}{\bibinfo{person}{Hye-Kyung Lee}.} \bibinfo{year}{2011}\natexlab{}.
\newblock \showarticletitle{Participatory media fandom: A case study of anime fansubbing}.
\newblock \bibinfo{journal}{\emph{Media, culture \& society}} \bibinfo{volume}{33}, \bibinfo{number}{8} (\bibinfo{year}{2011}), \bibinfo{pages}{1131--1147}.
\newblock


\bibitem[Liu et~al\mbox{.}(2021)]%
        {liu2021what}
\bibfield{author}{\bibinfo{person}{Xingyu Liu}, \bibinfo{person}{Patrick Carrington}, \bibinfo{person}{Xiang~'Anthony' Chen}, {and} \bibinfo{person}{Amy Pavel}.} \bibinfo{year}{2021}\natexlab{}.
\newblock \showarticletitle{What Makes Videos Accessible to Blind and Visually Impaired People?}. In \bibinfo{booktitle}{\emph{Proceedings of the 2021 CHI Conference on Human Factors in Computing Systems}}. \bibinfo{publisher}{ACM}, \bibinfo{address}{New York, NY, USA}, \bibinfo{pages}{1--4}.
\newblock


\bibitem[Liu et~al\mbox{.}(2022)]%
        {liu2022crossa11y}
\bibfield{author}{\bibinfo{person}{Xingyu Liu}, \bibinfo{person}{Ruolin Wang}, \bibinfo{person}{Dingzeyu Li}, \bibinfo{person}{Xiang'Anthony' Chen}, {and} \bibinfo{person}{Amy Pavel}.} \bibinfo{year}{UIST 2022}\natexlab{}.
\newblock \showarticletitle{CrossA11y: Identifying Video Accessibility Issues via Cross-modal Grounding}.
\newblock


\bibitem[Lu et~al\mbox{.}(2019)]%
        {lu2019vicariously}
\bibfield{author}{\bibinfo{person}{Zhicong Lu}, \bibinfo{person}{Michelle Annett}, {and} \bibinfo{person}{Daniel Wigdor}.} \bibinfo{year}{2019}\natexlab{}.
\newblock \showarticletitle{Vicariously experiencing it all without going outside: A study of outdoor livestreaming in China}.
\newblock \bibinfo{journal}{\emph{Proceedings of the ACM on Human-Computer Interaction}} \bibinfo{volume}{3}, \bibinfo{number}{CSCW} (\bibinfo{year}{2019}), \bibinfo{pages}{1--28}.
\newblock


\bibitem[McDaniel et~al\mbox{.}(2013)]%
        {mcdaniel2013evaluation}
\bibfield{author}{\bibinfo{person}{Troy McDaniel}, \bibinfo{person}{Lakshmie~Narayan Viswanathan}, {and} \bibinfo{person}{Sethuraman Panchanathan}.} \bibinfo{year}{2013}\natexlab{}.
\newblock \showarticletitle{An evaluation of haptic descriptions for audio described films for individuals who are blind}. In \bibinfo{booktitle}{\emph{2013 IEEE International Conference on Multimedia and Expo (ICME)}}. IEEE, \bibinfo{pages}{1--6}.
\newblock


\bibitem[Meta(2023)]%
        {fblive}
\bibfield{author}{\bibinfo{person}{Meta}.} \bibinfo{year}{2023 (accessed April 2023)}\natexlab{}.
\newblock \bibinfo{title}{Facebook Live}.
\newblock
\newblock
\newblock
\shownote{\url{https://www.facebook.com}}.


\bibitem[Microsoft(2023)]%
        {word365}
\bibfield{author}{\bibinfo{person}{Microsoft}.} \bibinfo{year}{2023}\natexlab{}.
\newblock \bibinfo{title}{Word for the web}.
\newblock
\newblock
\urldef\tempurl%
\url{https://www.microsoft365.com/launch/word}
\showURL{%
\tempurl}


\bibitem[Natalie et~al\mbox{.}(2020)]%
        {natalie2020viscene}
\bibfield{author}{\bibinfo{person}{Rosiana Natalie}, \bibinfo{person}{Ebrima Jarjue}, \bibinfo{person}{Hernisa Kacorri}, {and} \bibinfo{person}{Kotaro Hara}.} \bibinfo{year}{2020}\natexlab{}.
\newblock \showarticletitle{ViScene: A Collaborative Authoring Tool for Scene Descriptions in Videos}. In \bibinfo{booktitle}{\emph{The 22nd International ACM SIGACCESS Conference on Computers and Accessibility}}. \bibinfo{pages}{1--4}.
\newblock


\bibitem[Natalie et~al\mbox{.}(2023)]%
        {natalie2023supporting}
\bibfield{author}{\bibinfo{person}{Rosiana Natalie}, \bibinfo{person}{Joshua Tseng}, \bibinfo{person}{Hernisa Kacorri}, {and} \bibinfo{person}{Kotaro Hara}.} \bibinfo{year}{2023}\natexlab{}.
\newblock \showarticletitle{Supporting Novices Author Audio Descriptions via Automatic Feedback}. In \bibinfo{booktitle}{\emph{Proceedings of the 2023 CHI Conference on Human Factors in Computing Systems}}. \bibinfo{pages}{1--18}.
\newblock


\bibitem[of~the Blind(2003)]%
        {adpguidelines}
\bibfield{author}{\bibinfo{person}{American~Council of~the Blind}.} \bibinfo{year}{2003}\natexlab{}.
\newblock \bibinfo{title}{The Audio Description Project}.
\newblock
\newblock
\newblock
\shownote{\url{https://adp.acb.org/guidelines.html}}.


\bibitem[Pavel et~al\mbox{.}(2020)]%
        {pavel2020rescribe}
\bibfield{author}{\bibinfo{person}{Amy Pavel}, \bibinfo{person}{Gabriel Reyes}, {and} \bibinfo{person}{Jeffrey~P. Bigham}.} \bibinfo{year}{2020}\natexlab{}.
\newblock \showarticletitle{Rescribe: Authoring and Automatically Editing Audio Descriptions}. In \bibinfo{booktitle}{\emph{Proceedings of the 33rd Annual ACM Symposium on User Interface Software and Technology}} (Virtual Event, USA) \emph{(\bibinfo{series}{UIST '20})}. \bibinfo{publisher}{Association for Computing Machinery}, \bibinfo{address}{New York, NY, USA}, \bibinfo{pages}{747–759}.
\newblock
\showISBNx{9781450375146}
\urldef\tempurl%
\url{https://doi.org/10.1145/3379337.3415864}
\showDOI{\tempurl}


\bibitem[Peng et~al\mbox{.}(2021a)]%
        {peng2021slidecho}
\bibfield{author}{\bibinfo{person}{Yi-Hao Peng}, \bibinfo{person}{Jeffrey~P Bigham}, {and} \bibinfo{person}{Amy Pavel}.} \bibinfo{year}{2021}\natexlab{a}.
\newblock \showarticletitle{Slidecho: Flexible Non-Visual Exploration of Presentation Videos}. In \bibinfo{booktitle}{\emph{The 23rd International ACM SIGACCESS Conference on Computers and Accessibility}}. \bibinfo{pages}{1--12}.
\newblock


\bibitem[Peng et~al\mbox{.}(2021b)]%
        {peng2021say}
\bibfield{author}{\bibinfo{person}{Yi-Hao Peng}, \bibinfo{person}{JiWoong Jang}, \bibinfo{person}{Jeffrey~P Bigham}, {and} \bibinfo{person}{Amy Pavel}.} \bibinfo{year}{2021}\natexlab{b}.
\newblock \showarticletitle{Say It All: Feedback for Improving Non-Visual Presentation Accessibility}. In \bibinfo{booktitle}{\emph{Proceedings of the 2021 CHI Conference on Human Factors in Computing Systems}}. \bibinfo{pages}{1--12}.
\newblock


\bibitem[Project(2023)]%
        {obs}
\bibfield{author}{\bibinfo{person}{OBS Project}.} \bibinfo{year}{2023 (accessed May 2023)}\natexlab{}.
\newblock \bibinfo{title}{OBS: Open Broadcaster Software}.
\newblock
\newblock
\newblock
\shownote{\url{https://obsproject.com/}}.


\bibitem[Project(2019)]%
        {adp}
\bibfield{author}{\bibinfo{person}{The Audio~Description Project}.} \bibinfo{year}{2019}\natexlab{}.
\newblock \bibinfo{title}{adp.acb.org}.
\newblock
\newblock
\newblock
\shownote{\url{https://adp.acb.org/guidelines.html}}.


\bibitem[Reddit(2023)]%
        {blindsurveys}
\bibfield{author}{\bibinfo{person}{Reddit}.} \bibinfo{year}{2023 (accessed April 2023)}\natexlab{}.
\newblock \bibinfo{title}{r/BlindSurveys}.
\newblock
\newblock
\urldef\tempurl%
\url{https://reddit.com/r/blindsurveys}
\showURL{%
\tempurl}


\bibitem[S.A.(2023)]%
        {streamlabs}
\bibfield{author}{\bibinfo{person}{Logitech~Services S.A.}} \bibinfo{year}{2023 (accessed May 2023)}\natexlab{}.
\newblock \bibinfo{title}{Streamlabs}.
\newblock
\newblock
\newblock
\shownote{\url{https://streamlabs.com}}.


\bibitem[Sheng and Kairam(2020)]%
        {sheng2020virtual}
\bibfield{author}{\bibinfo{person}{Jeff~T Sheng} {and} \bibinfo{person}{Sanjay~R Kairam}.} \bibinfo{year}{2020}\natexlab{}.
\newblock \showarticletitle{From virtual strangers to irl friends: relationship development in livestreaming communities on twitch}.
\newblock \bibinfo{journal}{\emph{Proceedings of the ACM on Human-Computer Interaction}} \bibinfo{volume}{4}, \bibinfo{number}{CSCW2} (\bibinfo{year}{2020}), \bibinfo{pages}{1--34}.
\newblock


\bibitem[Smith et~al\mbox{.}(2013)]%
        {smith2013live}
\bibfield{author}{\bibinfo{person}{Thomas Smith}, \bibinfo{person}{Marianna Obrist}, {and} \bibinfo{person}{Peter Wright}.} \bibinfo{year}{2013}\natexlab{}.
\newblock \showarticletitle{Live-streaming changes the (video) game}. In \bibinfo{booktitle}{\emph{Proceedings of the 11th european conference on Interactive TV and video}}. \bibinfo{pages}{131--138}.
\newblock


\bibitem[Snyder(2005)]%
        {snyder2005audio}
\bibfield{author}{\bibinfo{person}{Joel Snyder}.} \bibinfo{year}{2005}\natexlab{}.
\newblock \showarticletitle{Audio description: The visual made verbal}. In \bibinfo{booktitle}{\emph{International Congress Series}}, Vol.~\bibinfo{volume}{1282}. Elsevier, \bibinfo{pages}{935--939}.
\newblock


\bibitem[Studios(2022)]%
        {incredibles}
\bibfield{author}{\bibinfo{person}{Pixar~Animation Studios}.} \bibinfo{year}{2004 (accessed August 2022)}\natexlab{}.
\newblock \bibinfo{title}{The Incredibles: Am I Fired Scene with Audio Description}.
\newblock
\newblock
\newblock
\shownote{\url{https://www.youtube.com/watch?t=128&v=2zhzVGmyjtg}}.


\bibitem[Twitch(2023)]%
        {twitch}
\bibfield{author}{\bibinfo{person}{Twitch}.} \bibinfo{year}{2023 (accessed April 2023)}\natexlab{}.
\newblock \bibinfo{title}{Twitch}.
\newblock
\newblock
\newblock
\shownote{\url{https://www.twitch.tv}}.


\bibitem[Walker and Nees(2011)]%
        {walker2011theory}
\bibfield{author}{\bibinfo{person}{Bruce~N Walker} {and} \bibinfo{person}{Michael~A Nees}.} \bibinfo{year}{2011}\natexlab{}.
\newblock \showarticletitle{Theory of sonification}.
\newblock \bibinfo{journal}{\emph{The sonification handbook}}  \bibinfo{volume}{1} (\bibinfo{year}{2011}), \bibinfo{pages}{9--39}.
\newblock


\bibitem[Wang et~al\mbox{.}(2021)]%
        {wang2021toward}
\bibfield{author}{\bibinfo{person}{Yujia Wang}, \bibinfo{person}{Wei Liang}, \bibinfo{person}{Haikun Huang}, \bibinfo{person}{Yongqi Zhang}, \bibinfo{person}{Dingzeyu Li}, {and} \bibinfo{person}{Lap-Fai Yu}.} \bibinfo{year}{CHI 2021}\natexlab{}.
\newblock \showarticletitle{Toward Automatic Audio Description Generation for Accessible Videos}.
\newblock


\bibitem[Yang et~al\mbox{.}(2022)]%
        {yang2022catchlive}
\bibfield{author}{\bibinfo{person}{Saelyne Yang}, \bibinfo{person}{Jisu Yim}, \bibinfo{person}{Juho Kim}, {and} \bibinfo{person}{Hijung~Valentina Shin}.} \bibinfo{year}{2022}\natexlab{}.
\newblock \showarticletitle{CatchLive: Real-Time Summarization of Live Streams with Stream Content and Interaction Data}. In \bibinfo{booktitle}{\emph{Proceedings of the 2022 CHI Conference on Human Factors in Computing Systems}} (New Orleans, LA, USA) \emph{(\bibinfo{series}{CHI '22})}. \bibinfo{publisher}{Association for Computing Machinery}, \bibinfo{address}{New York, NY, USA}, Article \bibinfo{articleno}{500}, \bibinfo{numpages}{20}~pages.
\newblock
\showISBNx{9781450391573}
\urldef\tempurl%
\url{https://doi.org/10.1145/3491102.3517461}
\showDOI{\tempurl}


\bibitem[YouTube(2023)]%
        {youtubelive}
\bibfield{author}{\bibinfo{person}{YouTube}.} \bibinfo{year}{2023 (accessed April 2023)}\natexlab{}.
\newblock \bibinfo{title}{YouTube Live}.
\newblock
\newblock
\urldef\tempurl%
\url{https://www.youtube.com/@live}
\showURL{%
\tempurl}


\bibitem[Yuan and Folmer(2008)]%
        {yuan2008blind}
\bibfield{author}{\bibinfo{person}{Bei Yuan} {and} \bibinfo{person}{Eelke Folmer}.} \bibinfo{year}{2008}\natexlab{}.
\newblock \showarticletitle{Blind hero: enabling guitar hero for the visually impaired}. In \bibinfo{booktitle}{\emph{Proceedings of the 10th international ACM SIGACCESS conference on Computers and accessibility}}. \bibinfo{pages}{169--176}.
\newblock


\bibitem[Yuksel et~al\mbox{.}(2020)]%
        {yuksel2020human}
\bibfield{author}{\bibinfo{person}{Beste~F Yuksel}, \bibinfo{person}{Pooyan Fazli}, \bibinfo{person}{Umang Mathur}, \bibinfo{person}{Vaishali Bisht}, \bibinfo{person}{Soo~Jung Kim}, \bibinfo{person}{Joshua~Junhee Lee}, \bibinfo{person}{Seung~Jung Jin}, \bibinfo{person}{Yue-Ting Siu}, \bibinfo{person}{Joshua~A Miele}, {and} \bibinfo{person}{Ilmi Yoon}.} \bibinfo{year}{2020}\natexlab{}.
\newblock \showarticletitle{Human-in-the-loop machine learning to increase video accessibility for visually impaired and blind users}. In \bibinfo{booktitle}{\emph{Proceedings of the 2020 ACM Designing Interactive Systems Conference}}. \bibinfo{pages}{47--60}.
\newblock


\end{thebibliography}

\appendix

\clearpage
\onecolumn
\appendix

\section{DESCRIBER PARTICIPANT DEMOGRAPHICS}

\begin{table*}[htp]
\small\sffamily
\def\arraystretch{1}
\setlength{\tabcolsep}{0.8em}

\centering
  
  \begin{tabular}{lccccccc}
    \toprule 
PID & Category           & Age & Gender         & Hours/Week   & Expertise & Audio & Text \\
    \midrule
P1  & Breath of the Wild & 30  & Male           & 15           & 8         & 1     & 5    \\
P2  & Chess              & 19  & Male           & 8            & 9         & 2     & 2    \\
P3  & Chess              & 21  & Male           & 3-4          & 7         & 3     & 3    \\
P4  & Chess              & 21  & Male           & 1            & 3         & 1     & 1    \\
P5  & Digital Art        & 21  & Female         & 1-2          & 9         & 1     & 2    \\
P6  & Digital Art        & 23  & Male           & 2            & 3         & 1     & 1    \\
P7  & Digital Art        & 23  & Female         & 0.5          & 8         & 1     & 3    \\
P8  & League of Legends  & 22  & Male           & 0-0.5        & 4         & 1     & 1    \\
P9  & League of Legends  & 19  & Male           & \textless{}1 & 9         & 1     & 1    \\
P10 & League of Legends  & 21  & Female         & $\sim$2      & 7         & 1     & 2    \\
P11 & Makeup             & 21  & Non-conforming & 30           & 9         & 3     & 3    \\
P12 & Makeup             & 21  & Female         & 0-2          & 2         & 1     & 1    \\
P13 & Smash Bros         & 21  & Male           & 6            & 10        & 1     & 1    \\
P14 & Smash Bros         & 21  & Male           & 10-12        & 9         & 1     & 1    \\
P15 & Smash Bros         & 22  & Male           & 1-2          & 9         & 3     & 6    \\
P16 & Valorant           & 21  & Female         & 6            & 8         & 1     & 1    \\
P17 & Valorant           & 21  & Male           & 5            & 6         & 2     & 8    \\
P18 & Valorant           & 21  & N/A            & 28           & 9         & 1     & 1  \\
\bottomrule
\end{tabular}
\caption{Demographics of describer study participants. ``Hours/Week'' refers to the number of hours per week the participant typically watches live video. ``Expertise'' refers to a self-reported, 1-10 scale of how familiar each participant felt with their chosen category. ``Audio'' and ``Text'' refer to a self-reported, 1-10 scale of how familiar each participant felt with writing audio or text descriptions before the study.}
\Description[Table depicting demographics of sighted describer participants]{Table 1: A table depicting the demographics of describer study participants. Participants are marked P1 through P18. Each participant is associated with a category, age, gender, hours per week, category expertise rating, audio description familiarity rating, and a text description familiarity rating. Participants are alphabetically assigned to categories Breath of the Wild, Chess, Digital Art, League of Legends, Makeup, Smash Bros, and Valorant. There is 1 Breath of the Wild participant, P1, two Makeup participants, P11 and P12, and the rest of the categories have three participants each. Ages range between 19 and 30, though over half the participants appear to have been 21. 11 participants are male, 5 are female, and 2 are non-conforming or N/A. Hours watched of video per week range on average from 0 to 30. Expertise in topic categories range from 2 to 10, with the majority being 7 or higher. Most audio and text description familiarities are listed as 1. No audio description familiarity is above 3, but 3 text description familiarity scores are 5 or higher.}
\label{tab:participantsdesc}
\end{table*}

\section{DESCRIPTION CODEBOOK}
\begin{table*}[htp]
\centering
\resizebox{\textwidth}{!}{%
\begin{tabular}{llccl}
\toprule
Code                 & Description                                                                & Quantity & Percentage & Example                                                                                                     \\
\midrule
Main:Play-by-Play    & “Minor” updates as they happen                                             & 215      & 71.67\%    & "She brushes setting powder under her eye left to right"                                                    \\
Main:State           & “Major” state updates (e.g. score updates, round changes, tempo shifts)    & 56       & 18.67\%    & "The new round starts"                                                                                      \\
\midrule
Action:Character     & Character-specific action names                                            & 44       & 14.67\%    & "Mythra Photon Edge"                                                                                        \\
Action:Verbiage      & Game-specific verbiage                                                     & 32       & 10.67\%    & "Sage plants spike"                                                                                         \\
Action:Controller    & Character-specific actions referenced using its controller input           & 5        & 1.67\%     & "Pyra whiffs an up-B"                                                                                       \\
\midrule
Camera:Streamer      & Focused on the streamer themselves                                         & 30       & 10\%       & "Frederic takes off his headphones"                                                                         \\
Camera:Background    & Focused behind the streamer or in their environment                        & 3        & 1\%        & "There are plant shelves in the background."                                                                \\
Camera:Misc          & From any additional cameras (e.g. dedicated dog camera)                    & 1        & 0.33\%     & "His dog looks up at him"                                                                                   \\
\midrule
Audio:Correction     & Correcting a previously recorded description (Audio only)                  & 11       & 3.67\%     & "No, not the Nexus. The Nexus towers."                                                                      \\
Audio:Unintelligible & Some part of the description is not understandable (Audio only)            & 2        & 0.67\%     & "Use another A ha{[}?{]} To catch uh Mew2King."                                                             \\
\midrule
Characters           & Character names                                                            & 105      & 35\%       & "Before picking Bard, we see his team has locked in Zeri ADC, Qiyana Jungle, Renekton top, Cassiopeia Mid." \\
Locations            & Use of words referring to locations specific to the media                  & 77       & 25.67\%    & "Timmy holds snowman and yellow from default on B site"                                                     \\
Lingo                & Use of words specific to the media or category of media                    & 32       & 10.67\%    & "Mew2King gets him with a landing neutral air forward tilt kill confirm."                                   \\
Object               & Use of objects, items, and/or utility                                      & 23       & 7.67\%     & “A prowler goes out as he nades mid"                                                                        \\
Tools                & Use or switching between of specific items to perform main content         & 18       & 6\%        & "She uses a brush to blend in a setting powder"                                                             \\
Uninformative        & Not able to be understood without additional context                       & 15       & 5\%        & "k"                                                                                                         \\
Text                 & On-screen text read audibly                                                & 11       & 3.67\%     & "His time ends with 27 correct and 3 incorrect"                                                             \\
Menu                 & Menus and inventory navigation; gamemode selection                         & 9        & 3\%        & "She goes into the inventory and switches the suit to the climbing gear."                                   \\
Commentary           & Side commentary from describer irrelevant to the stream content            & 8        & 2.67\%     & "ez takes out yasuo wannabe"                                                                                \\
Context              & Adds context about visual information to existing audio in the stream      & 5        & 1.67\%     & "White plays a move to support a pawn push which Alexandra doesn't think is necessary"                      \\
Overlays             & Content on streamer’s overlay                                              & 4        & 1.33\%     & "Subscriber emotes still bouncing across the screen"                                                        \\
Redundant            & Description that could have been discerned from audio alone                & 3        & 1\%        & "He wants to do some vision training" (After the streamer says they want to do some vision training)        \\
Chat                 & Describing messages from chat; giving context on streamer’s behalf & 1        & 0.33\%     & "There are a good amount of Fs in the chat."                                                               \\
\midrule
Miscellaneous        & Anything not marked in previous codes (with rationale)                     & 3        & 1\%        & "Fox pointed ears, long hair" (ambiguous) \\
\bottomrule
\end{tabular}%
}
\caption{List of codes applied to descriptions. Quantity and percentage metrics are out of 300 descriptions coded. Percentages add up to more than 100\% as multiple codes may be assigned for each description. In these descriptions, we have identified 4 higher-level themes: ``Main'', referring to the main content on stream, whether that be a game, the streamer's webcam, the streamer's drawing tablet; ``Action'', referring to different types of actions performed by characters in-game; ``Camera'', referring to descriptions of content based around one or more cameras in the scene; and ``Audio'', issues with descriptions as a result of the audio input method. Codes have been sorted in order of most prevalent to least prevalent, with higher-level themes grouped at the top, followed by individual lower-level themes, and miscellaneous at the bottom.\newline}
\Description[A table depicting various codes assigned to descriptions]{Table 2: A table depicting a codebook applied to descriptions. The table is divided into 5 sections of rows, being main, action, camera, audio, and miscellaneous. Each section is sorted in descending order by the quantity of descriptions assigned to each code. Each code includes a description of the code, a quantity of times it occurs, a percentage of the quantity divided by 300 total descriptions, and an example of a description that follows that code. The highest quantity code is Main:Play-by-Play with the description ``Minor'' updates as they happen, with 215 instances for 71.67\% of all descriptions coded. The corresponding example reads ``She brushes setting powder under her eye left to right.''}
\label{tab:codes}
\end{table*}

\section{AUDIENCE PARTICIPANT DEMOGRAPHICS}

\begin{table*}[!h]
    \small\sffamily
		\centering
\resizebox{\textwidth}{!}{%
  \begin{tabular}{lccccccccc}
   \toprule 
  PID & Age & Gender & Screenreader & Level of Vision             & \#1 Platform    & Hours/Week & Most-watched categories & Co-watched category & Chosen category \\
  \midrule
U1  & 35                                              & M      & VoiceOver    & Blind with light perception & YouTube         & 0.5-1      & Travel, Lifestyle     & Travel              & BOTW            \\
U2  & 30                                              & F      & VoiceOver    & Totally Blind               & YouTube         & 5-10       & Religious, Educational   & Lecture             & Art             \\
U3  & 29                                              & F      & \textbf{NVDA}, JAWS   & Blind with light perception & YouTube         & 0.25       & Music, Commentary     & Travel              & Chess           \\
U4  & 36                                              & M      & NVDA         & Blind with light perception &  \textbf{YouTube}, Twitch & 6-8      & Gaming                & Aviation            & BOTW            \\
U5  & 57                                              & M      & NVDA         & Totally Blind               & YouTube         & 1.5        & Commentary            & Lecture             & Smash           \\
U6  & 43                                              & F      & VoiceOver    & Totally Blind               & YouTube         & 1          & Gaming, Commentary    & Gaming              & BOTW            \\
U7  & 27                                              & M      & \textbf{NVDA}, JAWS   & Blind                       & YouTube         & 1          & Gaming                & Gaming              & BOTW            \\
U8  & 56                                              & M      & JAWS         & Blind with light perception & YouTube         & 5-6      & Music, Cooking, DIY   & Cooking             & Art             \\
U9  & 45                                              & M      & NVDA         & Totally Blind               & YouTube         & 0.75       & News, Gaming          & Gaming              & BOTW           \\
\bottomrule
\end{tabular}%
}
\caption{Demographics of user study participants. U3 and U7 cited familiarity with both NVDA and JAWS but opted to use NVDA for the study. U4 used YouTube and Twitch equally but opted to use YouTube for the co-watching portion of the study.}
\Description[A table depicting demographics of visually impaired participants]{Table 3: A table depicting demographics of user study participants marked U1 through U9. Each participant has an age, gender, screenreader, main platform they watch videos on, a category of video they co-watched with the researcher, and the chosen category they watched generated descriptions for. Ages range from 27 to 57, and there are 6 male participants alongside 3 female participants. 3 participants use voiceover, 5 use NVDA, and 1 used JAWS for the study. All participants had YouTube listed as the number one platform. The main co-watched category was gaming, with travel, lecture, aviation, and cooking also listed. 5 out of 9 participants chose to watch breath of the wild content, with 2 choosing art, 1 choosing smash, and 1 choosing chess.}
  \label{tab:participantsblv}
\end{table*}

\section{VIDEO REFERENCE}

\begin{table*}[htp]
\small\sffamily
\def\arraystretch{1}
\setlength{\tabcolsep}{0.8em}

\centering

  \begin{tabular}{lcccccc}
    \toprule 
Video ID & Category           & Streamer          & Intent             & On-screen chat & Cameras & Overlays \\
\midrule
V1       & Breath of the Wild & PointCrow         & Modified game      & x              & 1       & x        \\
V2       & Breath of the Wild & LimCube           & Speedrunning       & x              & 1       & x        \\
V3       & Breath of the Wild & DapperDame        & Let's Play         & x              & 1       & x        \\
V4       & Chess              & BotezLive         & Blitz              & x              & 1       & x        \\
V5       & Chess              & PrincepsComitatus & Classical          &                & 0       &          \\
V6       & Chess              & Keithonsky        & Puzzles            & x              & 1       & x        \\
V7       & Digital Art        & 39daph            & Stylized emojis    & x              & 1       & x        \\
V8       & Digital Art        & kaycem            & Educational        & x              & 2       & x        \\
V9       & Digital Art        & sezza             & Magical realism      & x              & 1       & x        \\
V10      & League of Legends  & loltyler1         & Mid-to-late game   & x              & 1       &          \\
V11      & League of Legends  & Doublelift        & Champion Select    & x              & 1       & x        \\
V12      & League of Legends  & LHS Esports       & Tournament Camera  &                & 0       &          \\
V13      & Makeup             & ThatGaymingAsian  & Creating a "look"  & x              & 1       & x        \\
V14      & Makeup             & CodeMiko          & Preparing for an event & x              & 2       &          \\
V15      & Makeup             & MeeshAmerica      & Preparing for work  &                & 1       &          \\
V16      & Smash Bros         & Hungrybox         & Elite Smash        & x              & 1       & x        \\
V17      & Smash Bros         & Mew2King          & Elite Smash        & x              & 1       & x        \\
V18      & Smash Bros         & Plup              & Elite Smash        & x              & 1       & x        \\
V19      & Valorant           & Masayoshi         & Ranked             & x              & 1       & x        \\
V20      & Valorant           & iiTzTimmy         & Deathmatch         & x              & 2       & x        \\
V21      & Valorant           & itzbelac          & Tournament Camera  &                & 0       &      \\
\bottomrule
\end{tabular}
\caption{Information about each video described. ``Intent'' refers to how a particular video is differentiated from other videos of the same category. ``On-screen chat'' contains an X if chat appeared on screen during the stream. ``Cameras'' refers to a number of webcams on-screen. ``Overlays'' contains an X if donations, subscription alerts, or on-screen text appears during the stream.}
\Description[A table depicting videos and their attributes]{Table 4: A table depicting video IDs V1 through V21. Each video has a corresponding Category, Streamer, Intent, whether or not it has on-screen chat, a number of cameras, and whether or not it has overlays. There are three videos for seven different categories. Each video is by a different streamer with a different intent. Most videos include on-screen chat, overlays, and at least one camera. Videos without any cameras also did not have any on-screen chat nor overlays.}
  \label{tab:livestreams}
\end{table*}

\end{document}